\documentclass[journal=ancac3,manuscript=article,layout=twocolumn]{achemso}

\usepackage[version=3]{mhchem} 

\author{Neerav Kharche}
 \email{kharcn@rpi.edu}
 \affiliation{Computational Center for Nanotechnology Innovations, Rensselaer Polytechnic Institute, Troy, NY 12180, USA}
 \altaffiliation{Department of Physics, Applied Physics and Astronomy, Rensselaer Polytechnic Institute, Troy, NY 12180, USA} 

 \author{Yu Zhou}
 \affiliation{Department of Physics, Applied Physics and Astronomy, Rensselaer Polytechnic Institute, Troy, NY 12180, USA}
 
 \author{Kevin P. O'Brien}
\affiliation{Intel Corporation Components Research, Hillsboro, OR 97124, USA}

\author{Swastik Kar}
 \affiliation{Department of Physics, Northeastern University, Boston, MA 02115, USA}

  \author{Saroj K. Nayak}
 \email{nayaks@rpi.edu}
 \affiliation{Department of Physics, Applied Physics and Astronomy, Rensselaer Polytechnic Institute, Troy, NY 12180, USA} 

\title[\texttt{achemso} demonstration]
{Effect of Layer-Stacking on the Electronic Structure of Graphene Nanoribbons}

\begin{document}
\begin{abstract}
The evolution of electronic structure of graphene nanoribbons (GNRs) as a function of the number of layers stacked together is investigated using \textit{ab initio} density functional theory (DFT) including interlayer van der Waals interactions. Multilayer armchair GNRs (AGNRs), similar to single-layer AGNRs, exhibit three classes of band gaps depending on their width. In zigzag GNRs (ZGNRs), the geometry relaxation resulting from interlayer interactions plays a crucial role in determining the magnetic polarization and the band structure. The antiferromagnetic (AF) interlayer coupling is more stable compared to the ferromagnetic (FM) interlayer coupling. ZGNRs with the AF in-layer and AF interlayer coupling have a finite band gap while ZGNRs with the FM in-layer and AF interlayer coupling do not have a band gap. The ground state of the bi-layer ZGNR is non-magnetic with a small but finite band gap. The magnetic ordering is less stable in multilayer ZGNRs compared to single-layer ZGNRs. The quasipartcle GW corrections are smaller for bilayer GNRs compared to single-layer GNRs because of the reduced Coulomb effects in bilayer GNRs compared to single-layer GNRs. \\\\
\end{abstract}

\section{Keywords}
graphene nanoribbons, electronic structure, GNR magnetism, graphene interconnects, quasiparticle band gaps.

\section{Introduction}
Graphene has attracted enormous attention due to its extraordinary electronic, optical, thermal and mechanical properties, and immense potentials for nanoelectronic applications \cite{Geim_NatureMaterials_2007,Novoselov_Science_2004,Novoselov_Nature_2005, Nair_Science_2008,Schedin_NatureMaterials_2007,Schwierz_NatureNano_2010,Balandin_NanoLetters_2008}. This single-atom thick, $sp^2$-hybridized allotrope of carbon  with a perfectly $2D$ confinement of its electronic states is a zero-gap semi-metal, exhibiting a linear dispersion relation $E(\textbf{k})$ = $\hbar v_F\textbf{k}$ near the meeting point of its conical valence and conduction bands (the Dirac point) \cite{Novoselov_Nature_2005}. Graphene can conduct much higher current densities than currently used Cu interconnects in the integrated circuits (ICs) \cite{Shao_APL_2008,Murali_APL_2009}. Graphene also exhibits much higher carrier mobilities compared to the conventional field-effect-transistor (FET) channel materials such as Si and III-Vs \cite{Novoselov_Science_2004}. These unique electronic properties make graphene one of the most promising candidate materials for both transistors and interconnects in future ICs \cite{ITRS_2009}.

For many practical applications, such as in nanoelectronic devices, graphene needs to be patterened into the so-called graphene nanoribbons (GNRs). Typically graphene is patterned by selectively removing the material by physical etching techniques. Graphene can also be patterned by selective chemical functionalization with Hydrogen \cite{Nieves_PRB_2010,Elias_Science_2009} or Fluorine \cite{Cheng_PRB_2010,Robinson_NanoLetters_2010}, which results in graphene nanostructures embedded in functionalized graphene: typically a wide bandgap insulator. The electronic properties of GNRs are very sensitive to their width and edge geometry \cite{Han_PRL_2007,Son_PRL_2006,Yang_PRL_2007,Son_Nature_2006}. The dependence of edge geometry on the electronic structure of GNRs has been mainly investigated using theoretical approaces such as the tight-binding model and density functional theory (DFT). Theoretical studies indicate that the band gap of single layer armchair GNRs (AGNRs) is extremely sensitive to their width \cite{Son_PRL_2006,Yang_PRL_2007}. Antiferromagnetic ordering at the edges of zigzag GNRs (ZGNRs) opens up a band gap while ferromagnetically ordered ZGNRs do not have a band gap \cite{Son_PRL_2006,Yang_PRL_2007}. The band gap in both AGNRs and ZGNRs can be controlled by an external electric field \cite{Son_Nature_2006}. Experimental observations of these theoretical predictions remain ilusive due to the challenging task of atomically precise control of the GNR edges. The available state-of-the-art transport measurements \cite{Han_PRL_2007,Shimizu_NatureNano_2011} are performed on GNRs with width on the order of $10 \ \rm nm$, however their edges are rough making direct quantitative comparison with theory difficult. The experimental techniques, however, keep on improving at a rapid pace. For example, the magnetic ordering at the edges predicted by DFT calculations \cite{Son_PRL_2006,Yang_PRL_2007} has been recently observed in GNRs with ultrasmooth edges \cite{Tao_NaturePhys_2011}.

Layer stacking provides an additional handle to tune the electronic properties of graphene and GNRs through interlayer interactions. The effect of layer stacking in graphene has been extensively studied theoretically \cite{Partoens_PRB_2006,Latil_PRL_2006,Partoens_PRB_2007,Grueneis_PRB_2008,Avetisyan_PRB_2010}. Recent experiments have demonstrated tuning of electronic properties of graphene by layer stacking. For example, (i) a band gap can be opened up in a bi-layer graphene by applying an electric field \cite{Zhang_Nature_2009}, (ii) tri-layer graphene behaves like a semimetal in the presence of an electric field \cite{Craciun_NatureNano_2009}, and (iii) multilayer graphene shows a peculiar conductivity spectra depending on the number of layers \cite{Mak_PNAS_2010, Wang_APL_2008}. There have been several theoretical studies on the electronic structure of bilayer GNRs \cite{Lam_APL_2008,Lima_PRB_2009,Lee_PRB_2005,Sahu_PRB_2008,Zhang_PRB_2010}, however, similar studies beyond bi-layer stacking are lacking except Ref. \cite{Sahu_PRB_2010}, where the effects of geometry relaxation were not included.

In this paper, we report an investigation of the effect of layer-stacking on the electronic structure of armchair and zigzag GNRs using \textit{ab initio} density functional theory (DFT). Interlayer van der Waals interactions are included to accurately model the effects of geometry relaxation on the electronic structure. The width and thickness dependence of the electronic structure  of multilayer AGNRs is discussed first, then the energetics of geometry relaxation and magnetic ordering and their effects on the electronic structure of multilayer ZGNRs are discussed.

It should be pointed out here that DFT based calculations underestimate the band gap of GNRs and perturbative correction schemes such as the GW method should be used to obtain accurate estimates \cite{Yang_PRL_2007}. GW calculations are computationally much more expensive compared to DFT. Since the primary goal here is to study the evolution of electronic structure as a function of the number of layers rather than quantitatively estimate the band gaps of stacked GNRs, we have used the computationally less expensive DFT scheme for most of the calculations. The more accurate GW corrected band gaps are calculated for single and bi-layer GNRs to provide an estimate of the quasiparticle corrections in these nanostructures.

\begin{figure}[t]
\resizebox{3.3in}{!}{
\includegraphics{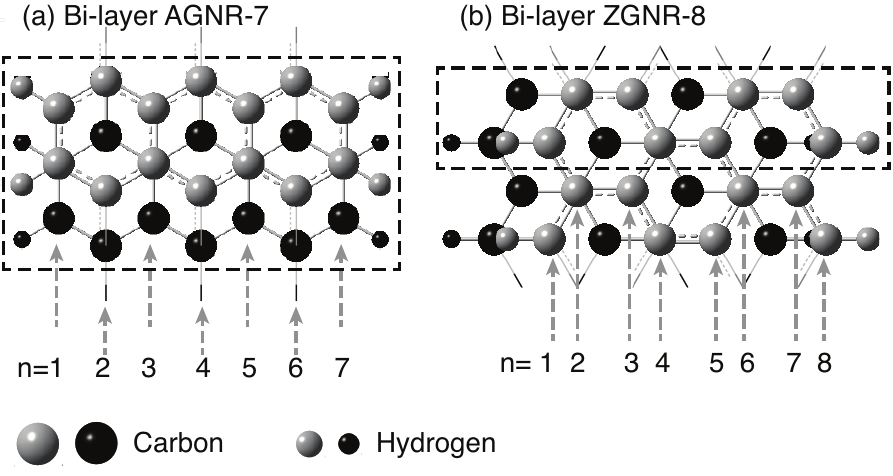}
}
\caption{Schematics of simulated structures. AB Bernal stacking in (a) AGNR-7 and (b) ZGNR-8. Grey and black colored atoms belong to top and bottom layers respectively. The unit cells are depicted by the dashed rectangles. }
\label{fig:AGNR_ZGNR_atoms}
\end{figure}

The atomistic schematics of simulated bi-layer AGNRs and ZGNRs with AB Bernal stacking are shown in ~\ref{fig:AGNR_ZGNR_atoms}(a) and ~\ref{fig:AGNR_ZGNR_atoms}(b) respectively. For three and more layers ABA and ABAB... stackings are used. An interlayer distance of $3.35$\, \AA\ (same as graphite) is used as a starting point in the geometry optimization. Edges of GNRs are passivated with hydrogen. Henceforth 1-8 layers thick GNRs are referred as few-layer GNRs while multilayer GNR refers to a GNR obtained by applying periodic boundary condition to a bilayer GNR in the thickness direction.

\section{Results and discussion}

\begin{figure}[t]
\resizebox{3.3in}{!}{
\includegraphics{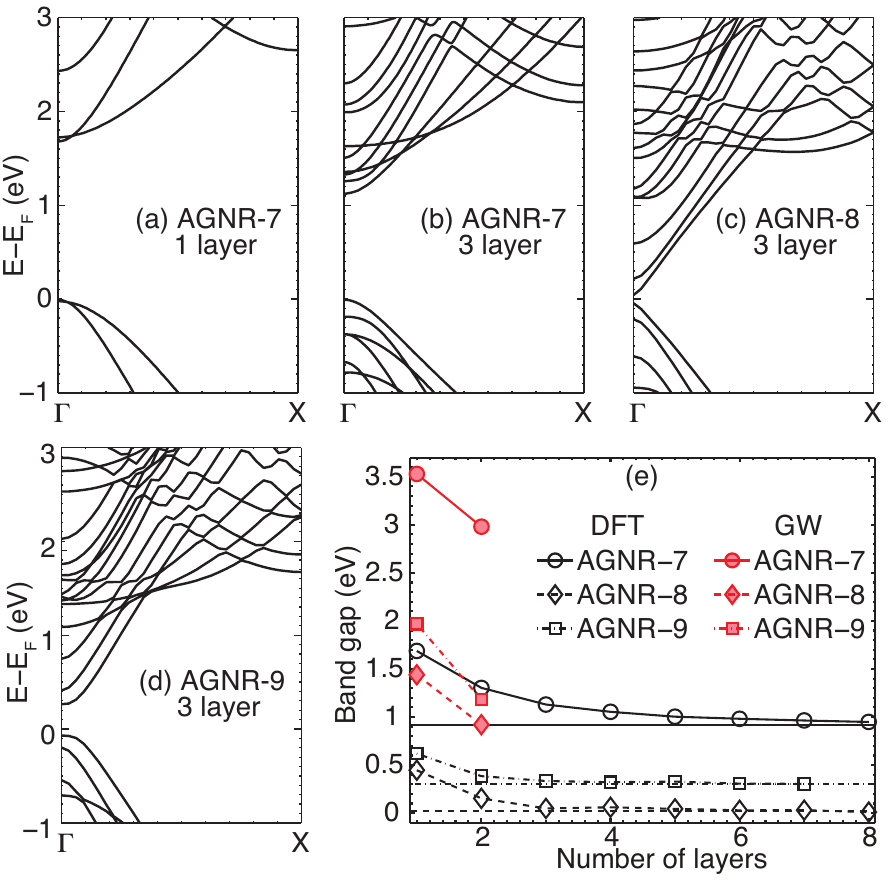}
}
\caption{Band structure of single- and tri-layer AGNRs belonging to three classes $3p$, $3p+1$, and $3p+2$ depending on their width. (a) Single-layer AGNR-7, (b) tri-layer AGNR-7, (c) tri-layer AGNR-8, and (d) tri-layer AGNR-9. (e) DFT band gaps (open symbols) of AGNRs as a function of the number of layers. The horizontal lines show the DFT band gaps of multilayer AGNRs obtained by applying periodic boundary conditions in the thickness direction. The solid symbols are the GW corrected band gaps of single- and bilayer AGNRs. }
\label{fig:AGNR_stack_bands_gaps}
\end{figure}

The electronic structure of AGNRs is discussed first. The band structure of a single-layer and a tri-layer AGNR-7 is shown in ~\ref{fig:AGNR_stack_bands_gaps}(a) and ~\ref{fig:AGNR_stack_bands_gaps}(b) respectively. As the three AGNR-7 layers are moved closer to form a tri-layer AGNR-7, the 3-fold band degeneracy is lifted due to the interlayer coupling and the band gap is reduced. Thus the degenerate bands in each of the three single-layer AGNR-7 results in three non-degenerate bands in a tri-layer AGNR-7. All few-layer AGNRs show a similar band structure evolution. As shown in ~\ref{fig:AGNR_stack_bands_gaps}(e), the band gap of AGNR-7 gradually decreases as the number of layers increases and shows a tendency to approach the multilayer AGNR-7 limit. The comparison of the band structures of tri-layer AGNR-7 (~\ref{fig:AGNR_stack_bands_gaps}(b)), AGNR-8 (~\ref{fig:AGNR_stack_bands_gaps}(c)), and AGNR-9 (~\ref{fig:AGNR_stack_bands_gaps}(d)) illustrates the fact that similar to single-layer AGNRs tri-layer AGNRs show three classes (\textit{i.e.}, $N=3p$, $3p+1$, and $3p+2$, where $N$ denotes the number of carbon chains along the width and $p$ is an integer) of band structures depending on their width. Similar to AGNR-7, AGNR-8 and AGNR-9 show convergence of the band gap to the multilayer limit (~\ref{fig:AGNR_stack_bands_gaps}(e)).  This classification was recently reported in Ref. \cite{Sahu_PRB_2010} for upto 4-layer thick AGNRs. ~\ref{fig:AGNR_stack_bands_gaps}(e) indicates that this classification persists beyond 4 layers.

The GW corrected band gaps of single- and bilayer AGNRs are also shown in ~\ref{fig:AGNR_stack_bands_gaps}(e). Large quasiparticle corrections on the order of $1-2 \ \rm eV$ are attributed to enhanced Coulomb effects \cite{Yang_PRL_2007}. Compared to bulk, GNRs are under confinement and have greatly reduced screening, which  enhance the Coulomb effects. The quasiparticle corrections for bilayer AGNRs are smaller compared to single-layer AGNRs. This is because each layer screens the Coulomb interactions in another layer reducing the overall Coulomb effects \cite{Neaton_PRL_2006,Thygesen_PRL_2009,Garcia-Lastra_PRB_2009,Freysoldt_PRL_2009,Li_JChemTheoComp_2009}.

ZGNRs show an edge magnetism, which can have multiple configurations depending on relative in-layer and interlayer spin polarizations \cite{Han_PRL_2007,Son_PRL_2006,Yang_PRL_2007,Son_Nature_2006,Lima_PRB_2009,Lee_PRB_2005,Sahu_PRB_2008,Zhang_PRB_2010,Sahu_PRB_2010}. The two possible spin polarizations in a single-layer ZGNR are (i) ferromagnetic (FM) in-layer and (ii) antiferromagnetic (AF) in-layer while multilayer ZGNRs can have several possible spin polarizations. The four spin polarizations in multilayer ZGNRs investigated here are (i) FM-FM: FM in-layer and FM interlayer, (ii) AF-FM: AF in-layer and FM interlayer, (iii) FM-AF: FM in-layer and AF interlayer, and (iv) AF-AF: AF in-layer and AF interlayer. Multilayer ZGNRs initialized in configurations (iii) and (iv) stay in those configurations while multilayer ZGNRs initialized in configurations (i) and (ii) converge to configuration (iv), which indicates that the FM interlayer coupling is not as stable as the AF interlayer coupling. Ref. \cite{Lee_PRB_2005} also reported that the AF interlayer coupling has lower energy compared to the FM interlayer coupling. Therefore, AF interlayer coupling is used in all ZGNR electronic structure calculations presented in this article.

\begin{figure}[t]
\resizebox{3.3in}{!}{

\includegraphics{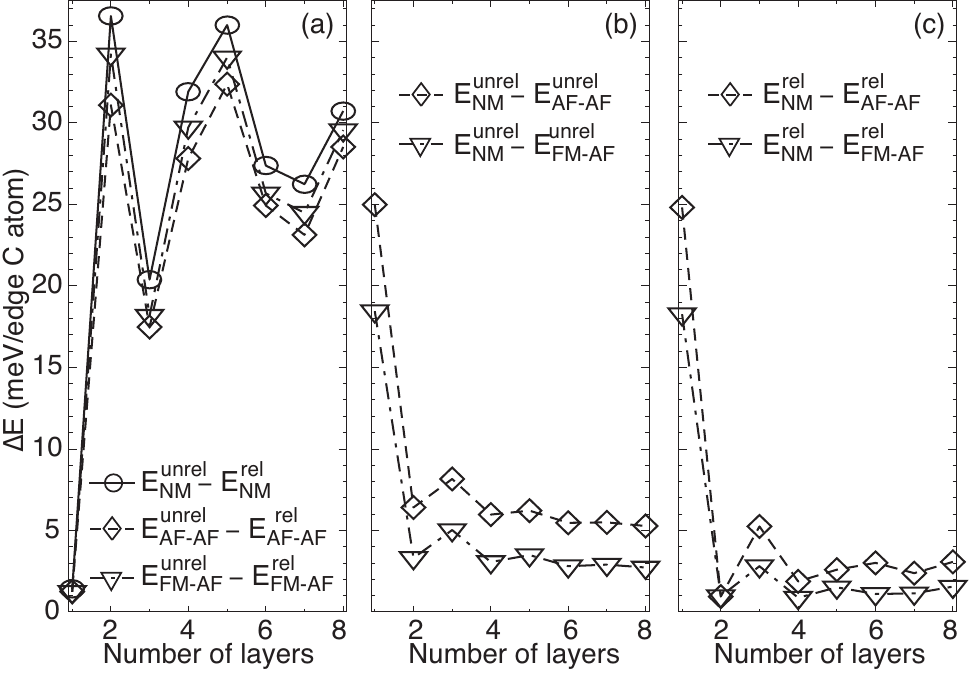}

}
\caption{Energetics of ZGNR-8. Difference between total energy of (a) relaxed and unrelaxed ZGNRs, (b) magnetically ordered and non-magnetic unrelaxed ZGNRs, and (c) magnetically ordered and non-magnetic relaxed ZGNRs as a function of the number of layers.}
\label{fig:ZGNR_energetics}
\end{figure}

The relaxation energy (\textit{i.e.} difference between total energy of relaxed and unrelaxed) ZGNRs as a function of the number of layers is plotted in ~\ref{fig:ZGNR_energetics}(a). Both non-magnetic (NM) and magnetically ordered ZGNRs show similar trends in the relaxation energy as a function of the number of layers. Due to the interlayer van der Waals attraction, the top and bottom layers show a concave curvature while the other layers remain more or less flat. Similar curvatures were reported in Ref. \cite{Lima_PRB_2009} for the bi-layer ZGNR. This concave curvature is the main cause of the reduction in total energy of the relaxed few-layer ZGNRs. The relaxation energy of the single-layer ZGNR is negligible because of the absence of curvature. 

~\ref{fig:ZGNR_energetics}(b) shows the magnetic stabilization energy ($\Delta E_M = E_{NM} - E_M$) of unrelaxed ZGNRs resulting only from the FM-AF and AF-AF orderings at the edges. ZGNRs with AF-AF and FM-AF orderings both show similar trends as a function of the number of layers such that ZGNRs with AF in-layer configuration are lower in energy compared to ZGNRs with FM in-layer coupling. $\Delta E_M$ is much smaller in ZGNRs with two or more layers compared to the single-layer ZGNR. This implies that the interlayer interations weaken the strength of magnetic ordering in each layer of the few-layer ZGNRs \cite{Lee_CPL_2004}. The interlayer interactions become stronger in relaxed ZGNRs because of the concave curvature in the top and bottom layers. This further weakens the strength of in-layer magnetic coupling, which results in lower energy difference between magnetic and non-magnetic states of relaxed ZGNRs compared to unrelaxed ZGNRs ~\ref{fig:ZGNR_energetics}(b,c). 

The bi-layer ZGNRs initialized in both the FM-AF and AF-AF configurations converge to a non-magnetic ground state when the geometry relaxation is allowed. Therefore in the bi-layer ZGNR, there is no reduction in the total energy due to the magnetic ordering ~\ref{fig:ZGNR_energetics}(c). Ref. \cite{Lima_PRB_2009}, which included van der Waals interactions also reported that the ground state of bi-layer ZGNRs is non-magnetic. In the calculations, where the geometry relaxation is not included, the bi-layer ZGNRs initialized in the FM-AF and AF-AF configurations stay in those configurations and do not converge to a non-magnetic ground state ~\ref{fig:ZGNR_energetics}(b). 

The calculated $\Delta E_M$ of relaxed ZGNRs (\ref{fig:ZGNR_energetics}(c)) is comparable to the thermal energy at room temperature ($k_B T \approx 25 \ \rm meV$), which indicates that magnetic ordering is unstable at room temperature. The layer stacking destabilizes edge magnetism by reducing $\Delta E_M$. The roughness and reconstruction at the ZGNR edges further destabilize the magnetic ordering even at low temperatures. Thus the magnetic ordering is expected to manifest at low temperatures in single-layer ZGNRs with atomically smooth edges. Indeed the edge magnetism was observed recently at low temperatures in GNRs with atomically smooth edges slightly offset from an ideal zig-zag edge \cite{Tao_NaturePhys_2011}.

\begin{figure}[t]
\resizebox{3.3in}{!}{
\includegraphics{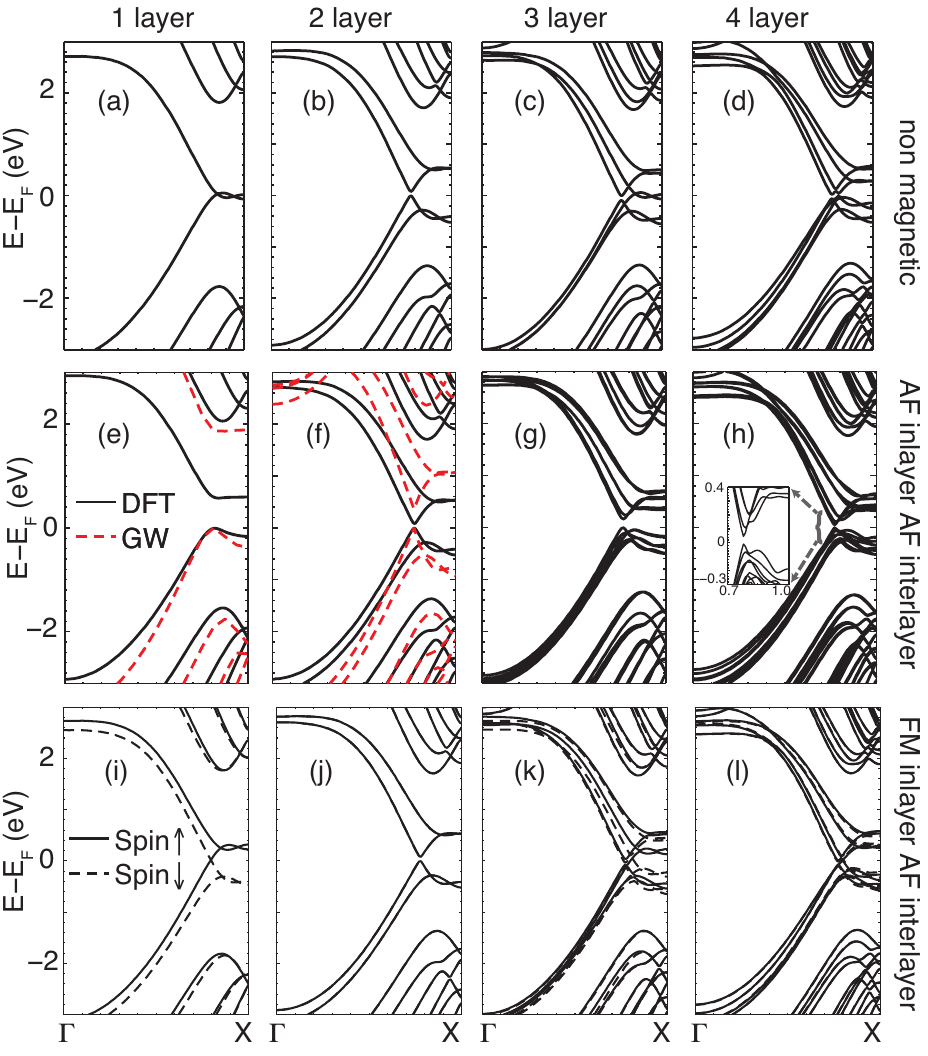}
}
\caption{DFT band structures of 1-4 layer ZGNR-8 in non-magnetic (a-d), antiferromagnetic (AF) in-layer and AF interlayer (e-h), and ferromagnetic (FM) in-layer and AF interlayer (i-l) configurations. Single-layer ZGNRs do not have AF interlayer couplings. Inset in (h) shows a zoomed in view of the band structure near the band gap. The dotted lines in (e) and (f) are the GW corrected band structures. The up and down spin states in (e-h) are degenerate while they are non-degenerate in (i-l). }
\label{fig:nonmag_AFAF_FMAF_1_4_bands}
\end{figure}

The band structures of 1-4 layer thick relaxed ZGNRs in the non-magnetic, AF-AF, and FM-AF configurations are shown in ~\ref{fig:nonmag_AFAF_FMAF_1_4_bands}. Non-magnetic ZGNRs have nearly flat bands near the Fermi level ($E_F$) leading to high density of states (DOS) near $E_F$. The bands near $E_F$ are mainly composed of the edge states. The high DOS near $E_F$ gives rise to magnetic instability and the edge states become spin-polarized \cite{Son_PRL_2006}. The up and down spin states in ZGNRs with AF-AF ordering at the edges are degenerate while they are split in ZGNRs with FM-AF ordering. The AF in-layer coupling at the edges opens up a band gap while FM in-layer coupling does not. The band structures of 5-8 layer thick relaxed ZGNRs are not shown here but are qualitatively similar. The band structures of the bi-layer ZGNR in the AF-AF (~\ref{fig:nonmag_AFAF_FMAF_1_4_bands}(f)) and FM-AF (~\ref{fig:nonmag_AFAF_FMAF_1_4_bands}(j)) configurations are identical to non-magnetic band structure (~\ref{fig:nonmag_AFAF_FMAF_1_4_bands}(b)) because  the bi-layer ZGNRs initialized in those configurations converge to the non-magnetic ground state after geometry relaxation. 

In the presence of the magnetic ordering at the edges, the non-magnetic bands in  ZGNRs are only slightly perturbed except the bands near $E_F$. Although the band structures of ZGNRs with even and odd number of layers look almost similar, they are different near $E_F$ and the Brillouin zone edge. Non-magnetic ZGNRs with odd number of layers have nearly flat bands crossing $E_F$ (~\ref{fig:nonmag_AFAF_FMAF_1_4_bands}(a,c)). Such bands are not present in non-magnetic ZGNRs with even number of layers. This property is reminiscent of the different electronic structure of graphene multilayers, where Dirac fermions with a linear dispersion are present in graphene with odd number of layers while only normal fermions with a parabolic dispersion are present in graphene with even number of layers \cite{Partoens_PRB_2007}. Upon magnetic ordering at the ZGNR edges, the splitting of the bands crossing $E_F$ is higher compared to the other flat bands slightly away from $E_F$. The bandstructures of ZGNRs with 4 or more layers are in close agreement with each other irrespective of their magnetic ordering (\ref{fig:nonmag_AFAF_FMAF_1_4_bands}(d,h,l)). This is also reflected in their total energies in ~\ref{fig:ZGNR_energetics}(c), where the energy difference between non-magnetic, AF-AF, and FM-AF orderings become very small as the number of layers increases beyond 3.

\begin{figure}[t]
\resizebox{3.3in}{!}{
\includegraphics[scale=1.0]{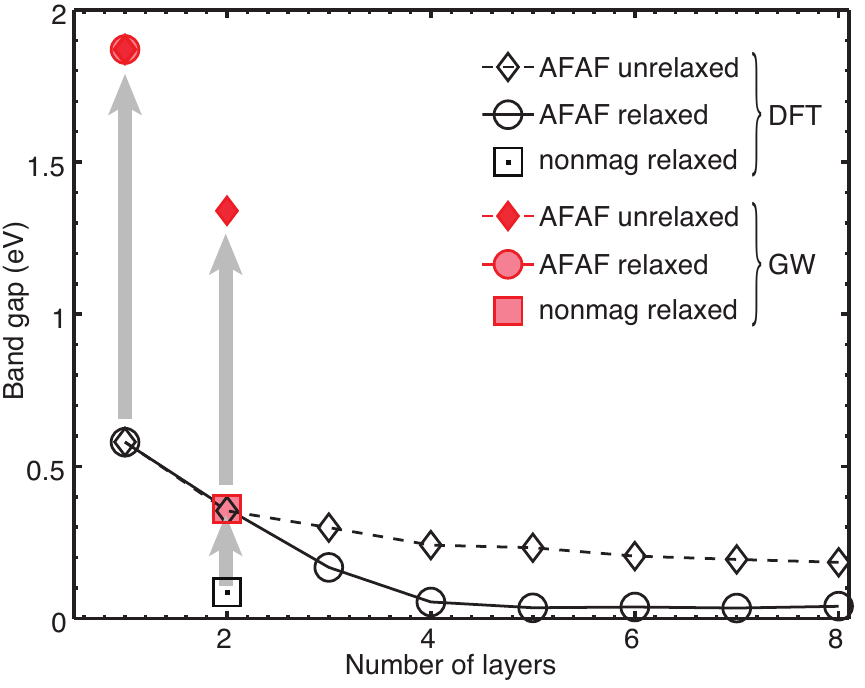}
}
\caption{DFT band gaps (open symbols) of relaxed and unrelaxed ZGNRs in AF-AF configuration as a function of the number of layers. The band gap of relaxed bi-layer ZGNR, which has a non-magnetic ground state, is depicted by an open square. The corresponding GW corrected band gaps of single- and bilayer ZGNRs are depicted by the solid symbols.}
\label{fig:AFAF_AFAF_flat_bandgap}
\end{figure}

The band gap of ZGNRs as a function of the number of layers is shown in ~\ref{fig:AFAF_AFAF_flat_bandgap}. As discussed earlier, only ZGNRs with AF-AF ordering and a non-magnetic bilayer ZGNR have a band gap. Relaxed ZGNRs have a smaller band gap compared to unrelaxed ZGNRs. The GW corrected band structures of a relaxed single-layer ZGNR with AF in-layer ordering and relaxed bi-layer ZGNR, which has a non-magnetic ground state are shown in ~\ref{fig:nonmag_AFAF_FMAF_1_4_bands}(e,f). The GW corrected band gaps of single- and bilayer ZGNRs are also shown in ~\ref{fig:AFAF_AFAF_flat_bandgap}. The quasiparticle corrections for unrelaxed bilayer AF-AF ZGNR are smaller compared to unrelaxed single-layer ZGNR with AF in-layer ordering because each layer screens the Coulomb interactions in another layer reducing the overall Coulomb effects \cite{Neaton_PRL_2006,Thygesen_PRL_2009,Garcia-Lastra_PRB_2009,Freysoldt_PRL_2009,Li_JChemTheoComp_2009}. Both the DFT and GW band gaps of a non-magnetic relaxed bilayer ZGNR are much smaller compared to unrelaxed bilayer ZGNR with AF-AF ordering.

Since the electronic structure of ZGNRs is mainly determined by the edge states \cite{Son_PRL_2006}, ZGNRs with different widths are expected to show similar electronic structure evolution with the number of layers as ZGNR-8 discussed here. Ultra-narrow ZGNRs may show different electronic structure evolution due to strong interedge interactions.

\section{Conclusion}
To summarize, the evolution of electronic properties of GNRs as a function of the number of layers stacked together is studied using DFT including van der Waals interactions. Multilayer AGNRs, similar to single-layer AGNRs, are found to exhibit three classes of band gaps depending on their width. In ZGNRs, the AF interlayer coupling is more stable compared to the FM interlayer coupling. ZGNRs with the FM in-layer and AF interlayer coupling do not have a band gap while ZGNRs with the AF in-layer and AF interlayer coupling have a finite band gap, which decreases with increasing the number of layers. The magnetic stabilization energy decreases as the number of layers increases indicating that the magnetic ordering is less stable in multilayer ZGNRs compared to single-layer ZGNRs. The ground state of the bi-layer ZGNR is found to be non-magnetic with a small but finite band gap.  The DFT calculations, which do not include geometry relaxation can not predict the non-magnetic ground state of a bi-layer ZGNR and overestimate the band gap of multilayer ZGNRs in AF-AF configuration. The GW calculations on single- and bilayer GNRs indicate that the quasiparticle band gap corrections decrease with increasing number of layers due to the reduction in Coulomb effects.

\section{Methods}
The electronic structure calculations are performed within the framework of first-principles density functional theory as implemented in the Vienna \textit{ab initio} simulation package (VASP) code \cite{kresse1996_VASP1,kresse1996_VASP2}. The PAW pseudopotentials \cite{Blochl_PRB_1994_PAW, Kresse_PRB_1999_VASP_PAW} and the PBE exchange correlation functional in the generalized gradient approximation \cite{Perdew_PRL_1996_PBE} are used. The DFT-D2 method of Grimme \cite{Grimme_JCC_2006} as implemented in VASP \cite{Bucko_JPCA_2010} is used to model the van der Waals (vdW) interaction between GNR layers. To ensure negligible interaction between periodic images, a large value ($10$\, \AA) of the vacuum region is used. The 1D Brillouin zone of few-layer GNRs is sampled using 32 uniformly spaced $k$-points while the 2D Brillouin zone of a multilayer GNR is sampled using 1$\times$16$\times$10 Monkhorst-Pack mesh \cite{Monkhorst_PRB_1976}. For the plane wave expansion of the wavefunction a $400$\, eV kinetic energy cut-off is used. The total energy and the atomic force are converged to within $10^{-4}$\, eV and $0.05$\, eV/\AA\ respectively. To obtain the band structure of few-layer GNRs a non-self-consistent calculation is carried out on $101$ uniformly spaced $k$-points in the positive half of the Brillouin zone using the converged charge density from the self-consistent calculation.

The GW calculations are performed using the ABINIT code \cite{Gonze_CompPhysComm_2009}. The norm-conserving pseudopotentials generated using the Trouiller-Martins scheme  \cite{Troullier_PRB_1991} implemented in the fhi98PP pseudopotential program \cite{Fuchs_CompPhysComm_1999} are used. The PBE parameterization for the exchange-correlation functional \cite{Perdew_PRL_1996_PBE} is used. To ensure negligible interaction between periodic images, a large value ($10$\,  \AA) of the vacuum region is used. The Brillouin zone is sampled using 32 uniformly spaced $k$-points. For the plane wave expansion of the wavefunction, a $12 \ \rm Ha$ kinetic energy cut-off is used. The DFT band structures calculated using VASP and ABINIT are virtually identical. The quasiparticle corrections are calculated within the $\rm G_0W_0$ approximation and the screening is calculated using the plasmon-pole model \cite{Hybertsen_PRB_1986}. The Coulomb cut-off technique proposed by S. I. Beigi \textit{et al.} \cite{Beigi_PRB_2006} is used to minimize the spurious interactions with periodic replicas of the system.

\acknowledgement
This work is supported partly by the Interconnect Focus Center funded by the MARCO program of SRC and State of New York, NSF PetaApps grant number 0749140. SK acknowledges financial support provided by NSF ECCS 1102481. Computing resources of the Computational Center for Nanotechnology Innovations at Rensselaer partly funded by State of New York and of nanoHUB.org funded by the National Science Foundation have been used for this work.


\providecommand*{\mcitethebibliography}{\thebibliography}
\csname @ifundefined\endcsname{endmcitethebibliography}
{\let\endmcitethebibliography\endthebibliography}{}


\end{document}